\documentclass[universe,article,accept,moreauthors,pdftex]{mdpi} 

\firstpage{1} 
\pubvolume{5}
\issuenum{5}
\articlenumber{117}
\pubyear{2019}
\copyrightyear{2019}
\history{Received: 17 April 2019; Accepted: 14 May 2019; Published: date}
\updates{yes} 

\pdfoutput=1

\Title{Quarkonium Production in the QGP}

\Author{Alexander Rothkopf \orcidA{}}

\AuthorNames{Alexander Rothkopf}

\address[1]{%
Faculty of Science and Technology, University of Stavanger, 4021 Stavanger, Norway; alexander.rothkopf@uis.no}

\abstract{We report on recent theory progress in understanding the production of heavy quarkonium in heavy-ion collisions based on the in-medium heavy-quark potential extracted from lattice QCD simulations. On the one hand, the proper in-medium potential allows us to study the spectral properties of heavy quarkonium in thermal equilibrium, from which we estimate the $\psi^\prime$ to $J/\psi$ ratio in heavy-ion collisions. On the other hand, the potential provides a central ingredient in the description of the real-time evolution of heavy-quarkonium formulated in the open-quantum-systems~framework.}


\begin{document}

\section{Introduction}

The bound states of heavy quarks and antiquarks, so-called heavy quarkonia, have matured into a high precision tool in heavy-ion collisions (HIC) at accelerator facilities, such as the Relativistic Heavy Ion Collider (RHIC) and the Large Hadron Collider (LHC). The availability of experimental data of unprecedented accuracy for both bottomonium $(b\bar{b})$ and charmonium $(c\bar{c})$, collected during the past five years, provides us access to different stages of the evolution of the quark--gluon plasma (QGP) created in the collision center. 

The STAR collaboration at RHIC has observed overall suppression of bottomonium states in $\sqrt{s_{NN}}=193$MeV collisions \cite{Adamczyk:2016dzv}. At LHC, the most recent dimuon measurements of the CMS collaboration at $\sqrt{s_{NN}}=5.02$TeV furthermore resolve a clear sign of excited state suppression \cite{Sirunyan:2017lzi}. These are compatible with phenomenological models that describe the bottom--anti-bottom pair as non-equilibrium test-particle traversing the QGP while sampling its full time evolution (see, e.g.,~\cite{Krouppa:2017jlg}). On the other hand, novel measurements by the ALICE collaboration have by now established an unambiguous signal for a finite elliptic flow, and even triangular flow of the charmonium vector channel ground state, the $J/\psi$ particle \cite{Acharya:2018pjd}. This tells us that the charm quarks must at least be in partial kinetic equilibrium with the bulk matter   to participate in its collective motion. In turn, equilibration entails a loss of memory of the initial conditions, positioning charmonium as probe of the late stages of the collision. 

The goal for theory thus must be to provide a first principles description of this intricate phenomenology. As the temperatures encountered in current heavy-ion collisions are relatively close to the chiral crossover transition, genuinely non-perturbative methods are called for and, in this article, I discuss one possible route how first principles lattice QCD simulations can contribute to gain insight into the equilibrium and non-equilibrium properties of heavy quarkonium in HIC.

\section{Quarkonium in Thermal Equilibrium}

Let us start with the question of what are the properties of heavy quarkonium in thermal equilibrium? That     is, we consider the idealized setting of immersing a heavy quark and antiquark pair in an infinitely extended QCD medium at a fixed temperature and wait until full kinetic equilibration is achieved. Then,   we ask for the presence or absence of in-medium bound eigenstates and their properties, such as their in-medium mass and stability. 

These questions may be answered in the modern language of quantum field theory by computing so-called in-medium meson spectral functions, which encode the particle properties as well defined peak structures. The position of the peaks along the frequency axis encodes the mass of the particle, while their width is directly related to the inverse lifetime of the state. At higher frequencies, the open heavy-flavor threshold manifests itself in the spectral function as broad continuous structures, often with a steep onset. In a thermal setting, the peak width   not only encodes the decay of the bound state into gluons but also carries a contribution from processes that: (1) excite the color singlet bound state into another singlet state due to thermal fluctuations; and (2) transform the singlet state into a color octet state due to the absorption of a medium gluon. On the level of the spectral function, these three contributions cannot be disentangled. Once we have access to the in-medium meson spectral function, we   argue that phenomenologically relevant processes, such as the production of $J/\psi$ particles at hadronization, may be estimated from inspecting the in-medium spectral structures.

There are currently two viable options to determine the in-medium quarkonium spectra in QCD and both involve lattice QCD simulations. For the first and direct one, we can compute the current--current correlators of a heavy meson in the Euclidean time domain, in which the simulation is carried out. In particular, for bottomonium, it is customary to use a discretization of the heavy quarks, which is derived from a non-relativistic effective field theory (EFT) (see,  e.g.,    \cite{Aarts:2014cda,Kim:2018yhk}). A fully relativistic description of bottomonium still requires too fine of a lattice spacing, which in turn would make simulation in dynamical QCD prohibitively expensive. For charmonium, relativistic formulations have been considered in, e.g., \cite{Ding:2012sp,Kelly:2018hsi}. From the Euclidean correlation function obtained in that way, the spectral function may be extracted using Bayesian inference. Due to the intricate structures encoded in the in-medium spectral function and the relatively small number of available simulated correlator points along the Euclidean time domain, this approach remains very challenging. Recent progress has been made in the robust determination of the ground state properties using the lattice NRQCD discretization at finite temperature \cite{Kim:2018yhk}. It was shown that the ground state of both   bottomonium and charmonium  becomes lighter as temperature increases. An investigation of the excited state properties however is currently still out of reach.

The second possibility is to take a detour and instead of the spectral function compute first the potential acting in between a static quark and antiquark at finite temperature. Using this in general complex valued potential one can solve a Schr\"odinger equation for the unequal time correlation function of meson color singlet wavefunctions, i.e. for the meson forward current--current correlator, whose imaginary part then yields the in-medium spectral function. This approach on the one hand provides us with a very precise determination of the spectral function, however it does not yet include finite velocity or spin dependent corrections, since only the static potential is used in the computation. At $T=0$, some of the correction terms to the heavy-quark potential have already been computed \cite{Koma:2007jq} and their determination at $T>0$ is a work in progress. We show below that,  to extract the in-medium potential from lattice QCD simulations,   a spectral function also needs to be reconstructed. However, the benefit here lies in the fact that the structure of this Wilson correlator spectral function is much simpler than that of the full in-medium meson spectral function and thus its reconstruction can be achieved with much higher precision. In this article, I focus on the second strategy.

Today we are in the fortunate position of not having to rely anymore on model potential for the description of heavy quarkonium. Indeed, over the past decade, it has become possible to derive the inter quark potential directly from QCD using a chain of EFTs \cite{Brambilla:2004jw}. An EFT provides a systematic prescription of how to exploit the inherent separation of scales between the heavy quark rest mass and the temperature, as well as the characteristic scale of quantum fluctuations in QCD $\Lambda_{\rm QCD}$ to simplify the language needed to describe the relevant physics of the in-medium two-body system. Starting out from the relativistic field theory QCD where heavy quarks are described by four-component Dirac spinors, one may go over to Non-Relativistic QCD (NRQCD), a theory of two-component Pauli spinors. Subsequently, we can leave the language of fermion fields all together and go over to an EFT called potential NRQCD (pNRQCD). The latter describes the quark antiquark pair in terms of color singlet and color octet wavefunctions with in general coupled equations of motions, containing both potential and non-potential effects. The potential in pNRQCD is nothing but a matching (Wilson) coefficient in the Lagrangian of the EFT.

It is the process of matching that allows us to connect back to QCD. We need to select a correlation function in the EFT and find the corresponding correlation function in QCD with the same physics content. Once we set them equal at the scale at which the EFT is supposed to reproduce the microscopic physics, we can express the non-local Wilson coefficients of the former in terms of correlation functions of QCD. For static quarks, it can be shown that the unequal time singlet wave function correlation function is related to the rectangular Wilson loop
\begin{align}
\langle \psi_s(t,r)\psi_s^*(0,r) \rangle_{pNRQCD} \overset{m\to\infty}{\equiv}W(t,r)=\left\langle  {\rm Tr} \left[ {\cal P} {\rm exp}\left( -ig \int dx^\mu A_\mu (x)\right)\right]\right\rangle_{QCD},
\end{align}
which obeys an equation of motion of the following type
\begin{align}
i\partial_t W(t,r)=\Phi(t,r)W(t,r)\quad \Phi(t,r)\in \mathbb{C}; \quad V(r)=\lim_{t\to\infty}\Phi(t,r)=\lim_{t\to\infty}\frac{i\partial_t W(t,r)}{W(t,r)} \label{Eq:DefPot}
\end{align}

If the function $\Phi$ at late times converges to a constant, we may use its value to define what we mean by the interquark potential \cite{Burnier:2012az}.

This genuinely real-time definition of the potential was first evaluated at high temperature in resummed perturbation theory by Laine et al. \cite{Laine:2006ns} who found it to be complex valued. The physics of the imaginary part has since been related to the phenomenon of Landau damping and gluo-dissociation~\cite{Beraudo:2007ky,Brambilla:2008cx}. Note that this complex potential does not evolve the wavefunction itself but instead a correlation function of wavefunctions. Thus, the presence of an imaginary part is by no means related to the disappearance of the heavy quarks (since they are static they cannot disappear from the system) but instead encodes the decoherence of the evolving in-medium system from its initial conditions \cite{Akamatsu:2011se}.

We may now ask how to evaluate the real-time definition of the potential in non-perturbative lattice QCD, as these simulations are carried out in artificial Euclidean time. It is here that the technical concept of spectral function again finds application \cite{Rothkopf:2009pk,Rothkopf:2011db}. Indeed, we may express the real-time Wilson line correlator $W(t,r)$ as a Fourier transform over its real-valued and positive definite spectral function $\rho(\omega ,r)$
\begin{align}
W(t,r)=\int d\omega e^{i\omega t} \rho(\omega,r) \quad \Leftrightarrow \quad W(\tau,r)=\int d\omega e^{-\omega \tau} \rho(\omega,r)
\end{align}

The quantity accessible on the lattice is the imaginary time Wilson correlator, which   is governed by the same spectral function, just with a different integral transform. Let me first note that using the spectral decomposition, inserted in the r.h.s. of Equation~\eqref{Eq:DefPot}, we can relate $\rho(\omega,r)$ and $V(r)$. A careful inspection of the relation between the two reveals that a potential picture is applicable as long as we can identify a well defined lowest lying peak structure in $\rho$ (for details see \cite{Burnier:2012az}). Its position is related to the real part of $V$, its width to the imaginary part. (In practice, we use the Wilson line correlators in Coulomb gauge instead of the Wilson loop in order to avoid the cusp divergences present in the latter.)

The central challenge lies in extracting the spectral function from lattice simulations, which amounts to solving an ill-posed inverse problem. In the past, this required the application of Bayesian inference \cite{Rothkopf:2019dzu}, which uses additional prior information available on the spectral function to regularize the inversion task. The benefit of the Bayesian strategy is that it is applicable to simulation data with moderate statistical uncertainty $(\Delta W/W \approx 10^{-2})$. One challenging aspect on the other hand is that the influence of the prior information on the end result must be carefully investigated. In this article, we   present the most recent results obtained for the potential using very high statistics simulations $(\Delta W/W<10^{-2})$. In that case, another method for spectral reconstruction becomes feasible, the Pade approximation \cite{Tripolt:2018xeo}. The simulation data are interpolated with an optimal Pade rational approximant, which in turn is analytically continued to give the retarded current--current propagator in real-time frequencies. Taking the imaginary part of this object yields the spectral function of interest, from which the values of ${\rm Re}[V]$ may be read off. Through mock datatests, we have found   that, based on $N_\tau=12,16$ data points, the Pade is yet unable to faithfully reconstruct the width of the potential peak, which is why we have resorted to extracting tentative values of ${\rm Im}[V]$ using standard methods of Bayesian inference \cite{Burnier:2013nla}. (For an alternative analysis based on the concept of effective potential, see~\cite{Petreczky:2017aiz}.)

The lattice data on which the latest determination of the potential is based were obtained in a collaboration with the HotQCD and TUMQCD collaboration \cite{Bazavov:2017dsy,Bazavov:2014pvz}. We compute the Wilson correlators on realistic $48^3\time12$ and $48^3\times 16$ lattices, featuring $N_f=2+1$ flavors of dynamical light quarks in the medium. These ensembles are deemed realistic, as the pion mass $m_\pi=161$ MeV lies close to its physical value. Temperature is changed via the lattice spacing in a large range of $T\in[151,1451]$ MeV.

\begin{figure}[H]
\centering
\begin{tabular}{cc}
\includegraphics[scale=0.4]{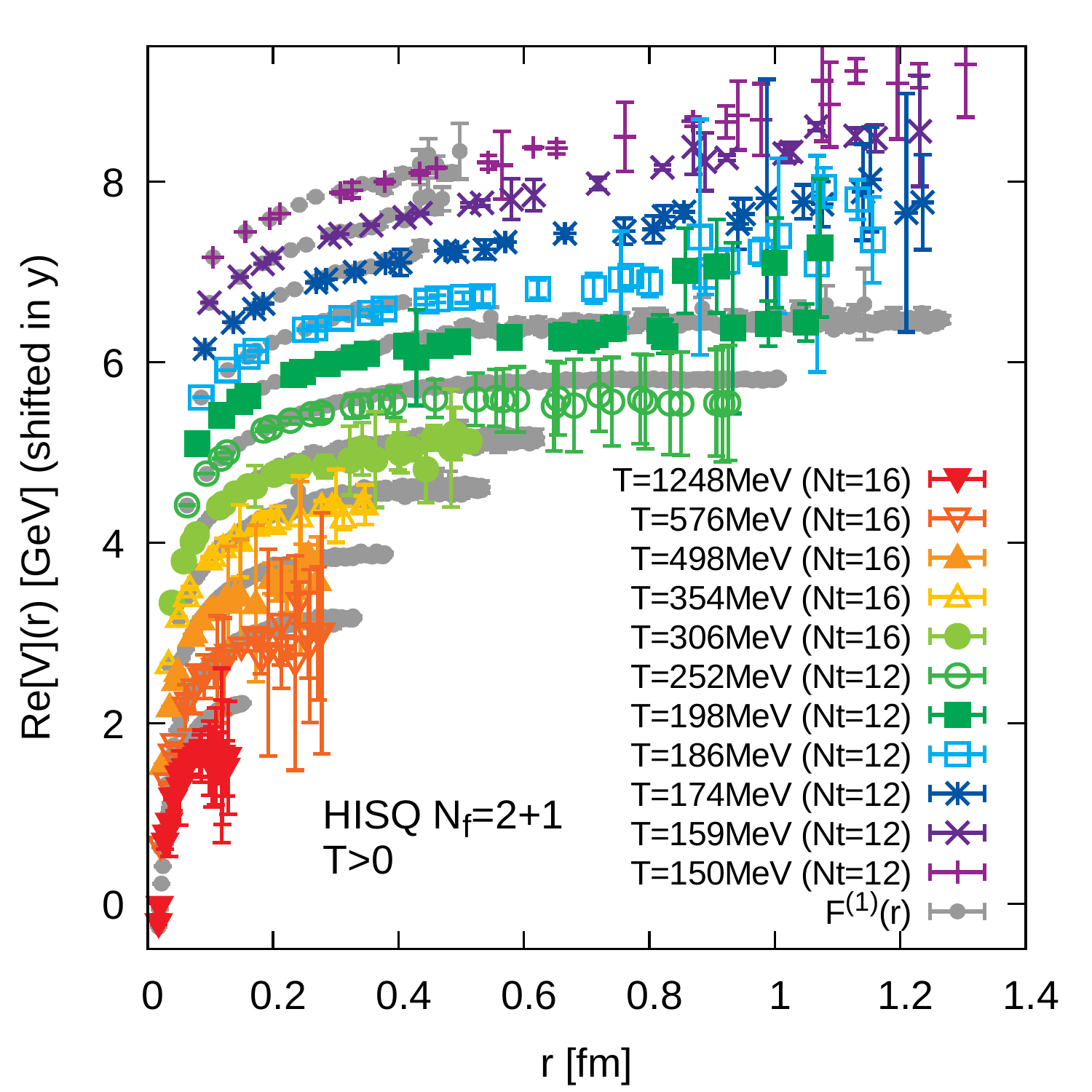}&
\includegraphics[scale=0.4]{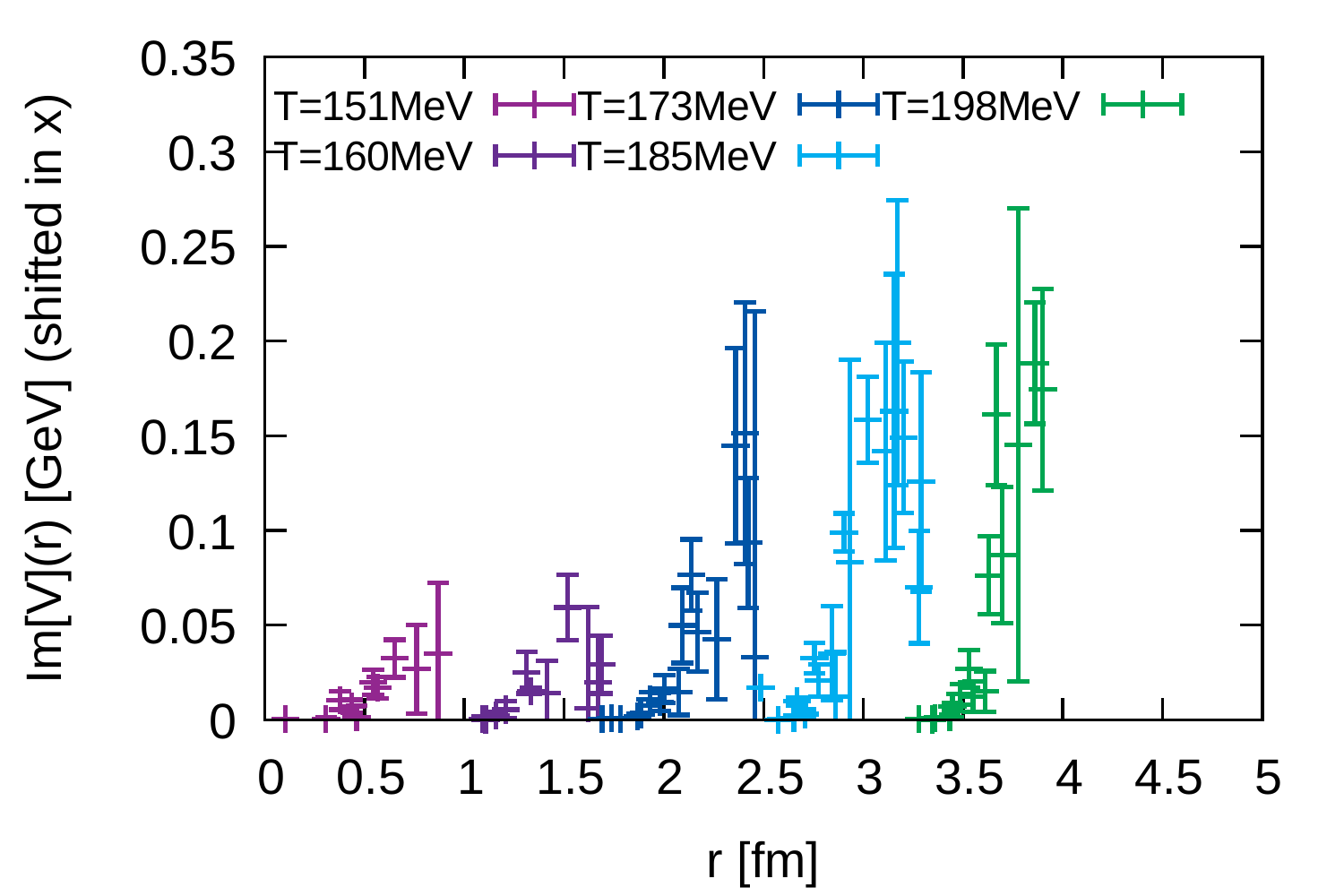}
\end{tabular}
\caption{( {\bf left}) ${\rm Re}[V]$ obtained from Pade reconstructed spectral functions of the Wilson line correlator in Coulomb gauge on $48^3\times 12,16$ lattices with $N_f=2+1$ light quarks. The values are shifted by hand in y-direction for better readability from lowest temperature $T=151$ MeV on top to highest $T=1451$~MeV bottom. The gray data points denote the color singlet free energy in Coulomb gauge on the same lattices. ({\bf right}) Tentative values of ${\rm Im}[V]$ at a selection of temperatures extracted via Bayesian inference from the same lattice data.}\label{Fig:LatPot}
\end{figure}

In  Figure \ref{Fig:LatPot} ({left}), we present the latest results for ${\rm Re}[V]$ from the aforementioned lattices \cite{Petreczky:2018xuh}. The~values shown are shifted manually in the y-direction for better readability. A qualitative inspection reveals that, while the potential in the hadronic phase at $T=151$ MeV is well described by a Cornell type potential (Coulombic at small distances, linear rising at larger distances), it quickly becomes weakened as one passes into the QGP regime. Well above $T=155$ MeV, ${\rm Re}[V]$ flattens off asymptotically and exhibits a form compatible with Debye screening.   Figure \ref{Fig:LatPot} ({right}) contains a selection of results for the imaginary part. As the extraction of spectral widths is much more challenging than that of the peak positions, the values shown are only tentative. (Since the Pade method is known to underestimate ${\rm Im}[V]$ based on $N_\tau=16$ data points, we show here results utilizing the Bayesian BR method instead.)

While it might be tempting to use the lattice values of ${\rm Re}[V]$ and ${\rm Im}[V]$ directly for a subsequent computation of the in-medium spectral function, this is not admissible. The lattice results obtained here are not yet extrapolated to the continuum limit and thus will not lead to consistent phenomenological results. Obtaining a genuine extrapolation is a work in progress but has thus far not yet been achieved. Therefore, continuum corrections need to be used as laid out in detail, e.g., in Ref.~\cite{Burnier:2015tda}.    To utilize the discrete values of the in-medium potential to solve a Schr\"odinger equation requires an analytic parametrization of ${\rm Re}[V]$ and ${\rm Im}[V]$ that can faithfully reproduce the lattice data. A novel derivation of such a parameterization \cite{Lafferty:2019jpr}, based on the generalized Gauss law, has been presented at the 2018 Zimanyi workshop (see Ref.~\cite{LaffertyZimanyi}).

\begin{figure}[H]
\hspace{-1cm}
\begin{tabular}{cc}
\includegraphics[scale=0.45]{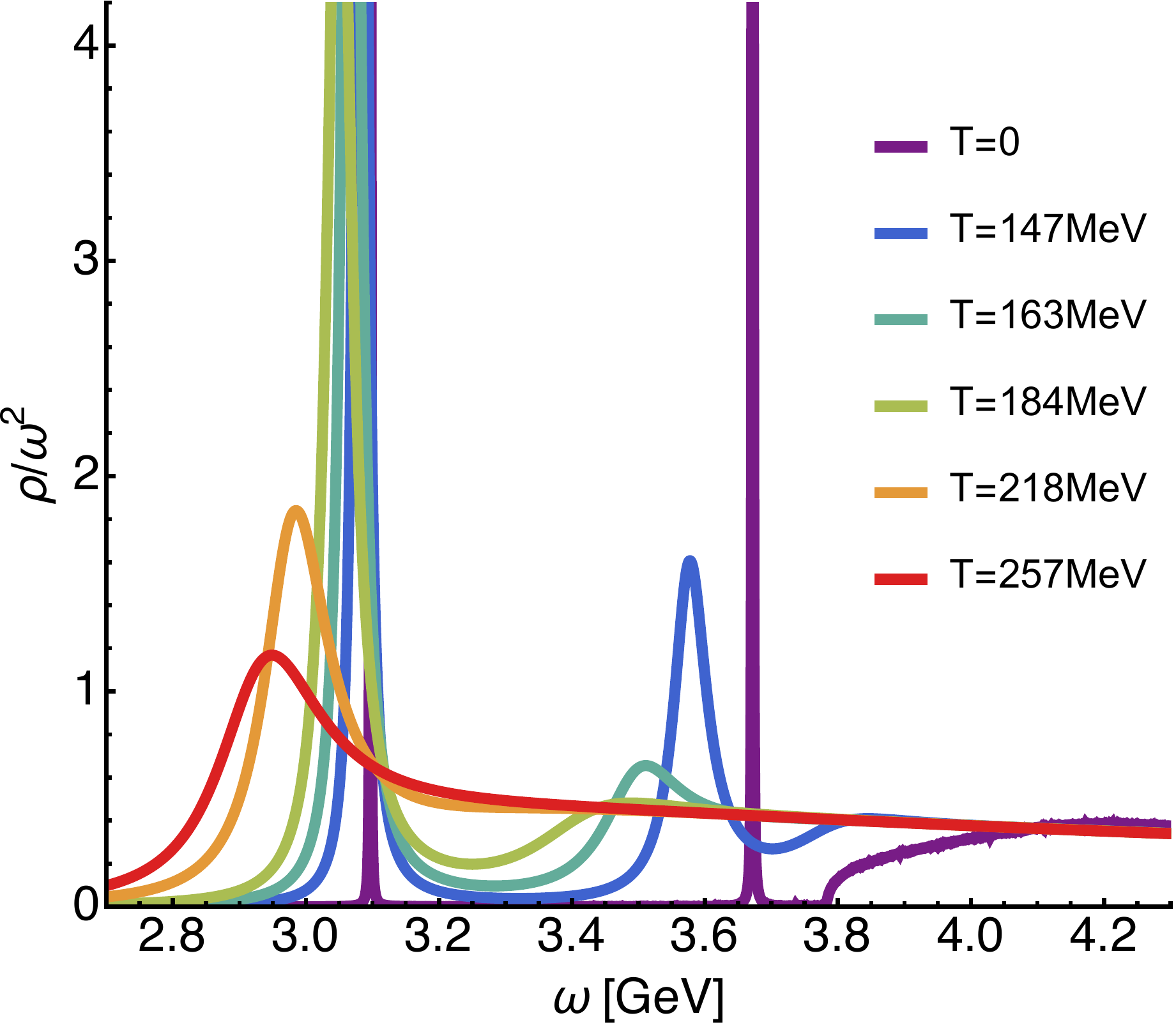}&
\includegraphics[scale=0.6]{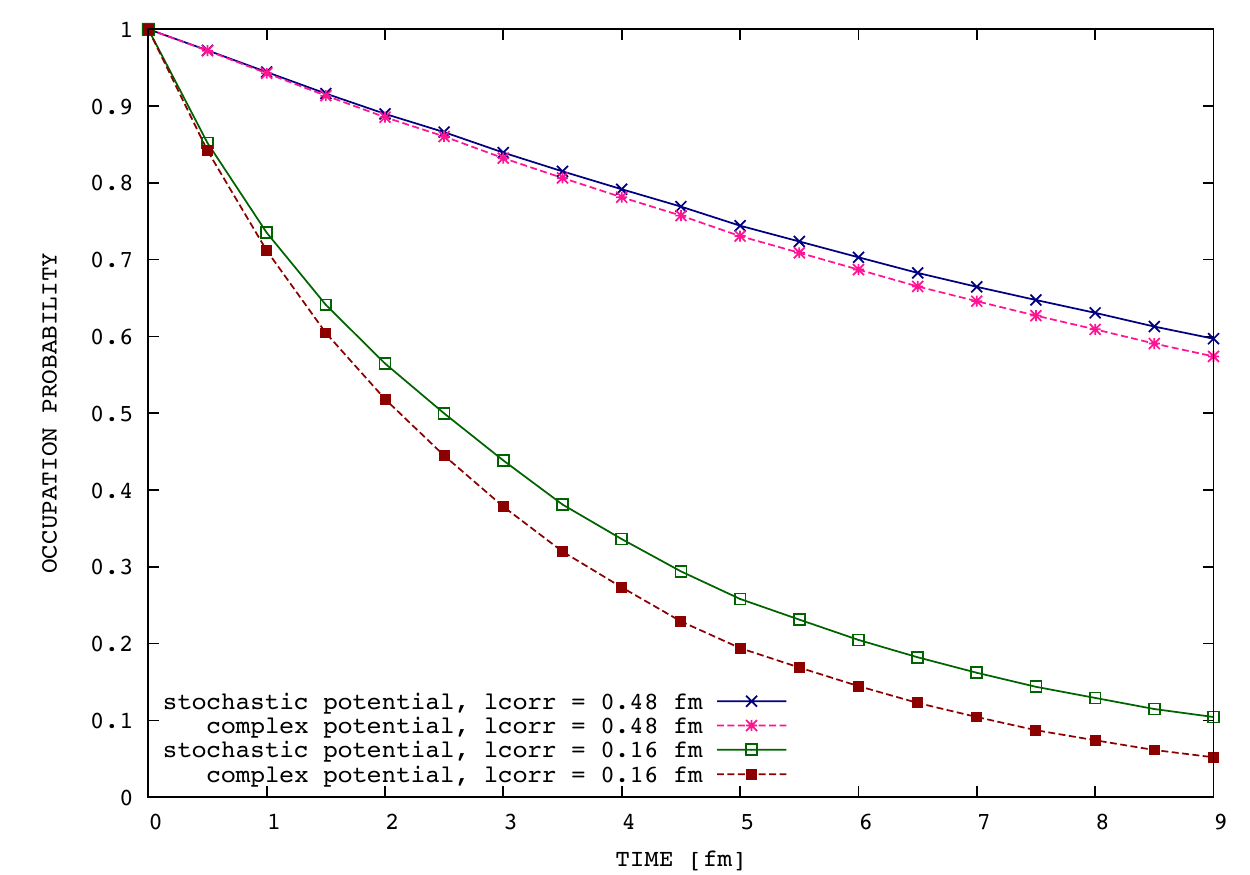}
\end{tabular}
\caption{({\bf left}) Charmonium in-medium spectral functions from the continuum corrected in-medium heavy quark potential \cite{Burnier:2015tda}. ({\bf right}) Survival probabilities of the ground state in a one-dimensional model calculation of the real-time dynamics of bottomonium in the open-quantum systems approach~\cite{Kajimoto:2017rel}. The blue and green curve correspond to the stochastic potential computation with different correlation lengths. The pink and dark red curves arise from a naive Schr\"odinger equation with complex potential.}\label{Fig:InmedRes}
\end{figure}

In  Figure \ref{Fig:InmedRes} (left), we show the in-medium spectral functions computed from the continuum corrected in-medium heavy quark potential obtained in \cite{Burnier:2015tda}. One can clearly see the characteristic in-medium modification consisting of a shift of the peaks to lower frequencies and a concurrent broadening before they are dissolved into the continuum structure, whose onset moves to lower and lower frequencies. Consistent with intuition, the more weakly bound excited state is more strongly affected by the medium than the deeply bound ground state.

How can such spectral functions help us to learn about quarkonium production in HICs? Note that we are considering a fully thermalized scenario here, which applies, if at all, for charmonium. Note further that what is measured in experiment are not the decay dileptons from the in-medium states but the decays of vacuum states long after the QGP ceases to exist. Thus, any information of in-medium quarkonium needs to be translated into a modification of the yields of produced vacuum states at hadronization. The process of hadronization is among the least well known stages of a HIC and a first principles understanding of its dynamics has thus  far not been achieved. Therefore, we continue with the phenomenological ansatz of instantaneous freezeout introduced in \cite{Burnier:2015tda}. That     is, we assume that at the phase boundary the in-medium states convert into vacuum states. The question we then wish to answer is: How many vacuum states does the in-medium spectral peak correspond to? The answer may be given in units of dilepton emission $ R_{\ell\bar\ell}\propto \int dp_0 d^3\mathbf p \frac{\rho(P)}{P^2}n_B(p_0)$, which relates to the area under the spectral peaks.

That     is, we compute the weighted area under the in-medium $J/\psi$ peak and divide by the area of the vacuum spectral peak. This is our estimate for the number of $J/\psi$ particles produced in this scenario. Carrying out the same computation for the $\psi^\prime$ peak, we may form the ratio of the two results, which constitutes our estimate for the in-medium $\psi^\prime$ to $J/\psi$ ratio. The value obtained in Ref. \cite{Burnier:2015tda} reads
\begin{align}
R_{\ell\bar\ell}^{\psi'}/ R_{\ell\bar\ell}^{J/\psi}=0.023\pm0.004.
\end{align}
and agrees within uncertainty with the value predicted by the statistical model of hadronization \cite{Andronic:2009sv}, as well as with the most recent determination of the ratio by the ALICE collaboration at the LHC (see,  e.g.,    \cite{Andronic:2018vqh}).

\section{In-Medium Quarkonium Real-Time Dynamics}

Up to this point, we have only considered equilibrium aspects of quarkonium. In a HIC, this will always constitute only an approximation to the genuine non-equilibrium physics occurring. Therefore, we wish to learn more about the real-time dynamics of quarkonium states exploiting the fact that we already have access to the in-medium potential extracted on the lattice. A promising route towards a microscopic understanding of quarkonium real-time dynamics is offered by the open-quantum systems approach, a technique developed originally in the context of condensed matter theory. 

The overall system consisting of the heavy quark and antiquark, as well as the medium degrees of freedom is of course closed and described by a hermitean Hamiltonian. The overall density matrix evolves according to the von Neumann equation
\begin{align}
H=H_{Q\bar Q}\otimes I_{med}+I_{Q\bar Q}\otimes H_{med}+H_{int}, \quad \frac{d\rho}{dt}=-i[H,\rho].
\end{align}

Our goal however is to investigate the properties and dynamics of the heavy quarkonium sub system coupled to the thermal bath. To this end, we may trace out all medium degrees of freedom from the density matrix of the full system, ending up with  $\rho_{Q\bar Q}={\rm Tr_{med}}[\rho]$. The question then is: What kind of equation of motion does this reduced density matrix obey?

Over the past five years, it has become possible to derive the master equation for $\rho_{Q\bar Q}$ from QCD, based on a limited number of assumptions \cite{Akamatsu:2012vt,Akamatsu:2014qsa}. Starting from the path integral representation of the density matrix on the Schwinger--Keldysh contour, the integrating out of the medium degrees of freedom may be implemented in a functional sense. This leads to a path integral for the reduced density matrix, in which only the heavy quark degrees of freedom appear explicitly. In addition to the heavy quark action on the forward and backward contour, an additional effective action emerges, the so-called Feynman--Vernon influence functional $S_{FV}$. It encodes all interactions between the subsystem and the traced out medium. $S_{FV}$ in general is a very complicated object but it may be simplified using the separation of scales in the system. As shown in Ref.~\cite{Akamatsu:2014qsa}, at high temperatures, where at intermediate steps of the derivation a weak coupling ansatz has been used, the Feynman--Vernon influence functional takes the explicit form
\begin{align}
S_{FV}\approx S_{pot}\big[{\rm Re}[V]\big]+S_{fluct}\big[{\rm Im}[V]\big] + S_{diss}\big[{\rm Im}[V]\big] + S_{LB}.
\end{align}

The first part is related to a real valued in-medium potential term, while the second and third implement the fluctuation--dissipation relation for the heavy quarkonium. They are intimately related to the imaginary part of the interquark potential. The last term assures that the master equation for $\rho_{Q\bar Q}$ preserves the positivity of its eigenvalues. (For other recent studies of the open-quantum systems approach for quarkonium, see \cite{DeBoni:2017ocl,Blaizot:2017ypk,Brambilla:2017zei,Yao:2018nmy,Brambilla:2019tpt}.)

The above expression for $S_{FV}$ leads to Markovian dynamics for $\rho_{Q\bar Q}$, described by a so called Lindblad equation.
\begin{align}
\frac{d}{dt} \rho_{Q\bar{Q}}(t)=-i\big[ H_{Q\bar{Q}}, \rho_{Q\bar{Q}}\big] + \sum_{i=1}^{N_{LB}}\gamma_i\Big( \hat L_i \rho_{Q\bar{Q}} \hat L_i^\dagger - \frac{1}{2} \hat L_i \hat L_i^\dagger \rho_{Q\bar{Q}} - \frac{1}{2} \rho_{Q\bar{Q}} \hat L_i \hat L_i^\dagger \Big)
\end{align}

The operators $L_i$ are called Lindblad operators and encode the interactions between the quarkonium subsystem and the surrounding environment. They may be expressed in terms ${\rm Im}[V]$. It is important to note that the Lindblad equation cannot be implemented (unraveled) in terms of a deterministic evolution of a microscopic wave function. Instead, one is led to stochastic dynamics for an ensemble of wavefunctions.

Together with collaborators from Japan, we have investigated the effects of the Lindblad operators on the real-time dynamics of heavy quarkonium in a simple one-dimensional setting \cite{Kajimoto:2017rel}. As a first step, we considered only the leading order gradient expansion of $S_{FV}\approx S_{pot}\big[{\rm Re}[V]\big]+S_{fluct}\big[{\rm Im}[V]\big]$, which leads to the notion of a stochastic potential. That     is,    it allows   implementing unitary time evolution via ${\rm Re}[V]$, which is stochastically disturbed with noise $\eta$, whose correlations are governed by ${\rm Im}[V]$.

\begin{align}
\psi_{Q\bar{Q}}(t)={\rm exp}\Big[ -\frac{\nabla^2}{M}+{\rm Re}[V]+\eta(t) \Big]\psi_{Q\bar{Q}}(0), \quad i\partial_t \langle \psi_{Q\bar{Q}}(t)\rangle =\Big( -\frac{\nabla^2}{M}+{\rm Re}[V] - i |{\rm Im}[V]| \Big)\langle \psi_{Q\bar{Q}}(t)\rangle
\end{align}

While the evolution of each realization of the ensemble proceeds via a norm preserving evolution operator, the ensemble average of the wave function washes out according to a Schr\"odinger equation with a complex valued potential. This mechanism provides a unitary microscopic implementation of quarkonium real-time dynamics, which reproduces the imaginary part of the interquark potential for the unequal time correlation function of wavefunctions.

Note that there is a new physical scale present in this approach, which is the correlation length of the noise induced by the medium. Depending on the size of the quarkonium bound state compared to this correlation length, the noise may be able to efficiently destabilize the bound state or not. This phenomenon is known as decoherence. That     is,    the noise provides an additional mechanism to dissociate a heavy quarkonium particle over time, which acts in addition to the screening of the real-valued potential. 

In Figure \ref{Fig:InmedRes} (right), we show an example computation of the survival probabilities of the bottomonium ground state in a one dimensional setup based on perturbative values for the in-medium ${\rm Re}[V]$ and ${\rm Im}[V]$. We draw two conclusions. First, the survival crucially depends on the value of the medium correlation length. Secondly, using the more realistic description in terms of a stochastic potential instead of a naive Schr\"odinger equation with a complex potential leads to significantly different survival. The naive approach systematically underestimates the survival. 

While the stochastic potential provides a conceptually attractive microscopic implementation of the complex inter-quark potential, it can only be the first step towards understanding heavy quarkonium in-medium dynamics. It does not account for dissipation effects and thus does not allow the quarkonium to thermalize with its surroundings. This means that the stochastic potential description is only applicable to early times in the evolution. Incorporation of the full Linblad equation is work in progress and we have successfully tested it in the single heavy quark case \cite{Akamatsu:2018xim}. The extension to quarkonium is under way.

\section{Summary}

In this article, I have showcased recent progress in our understanding of in-medium heavy quarkonium in the context of heavy-ion collisions. In thermal equilibrium, it has become possible to derive a complex valued real-time in-medium potential from QCD based on EFT methods. Its~evaluation in lattice QCD simulations is challenging as it involves the reconstruction of spectral functions from Wilson correlators. The most recent determination has been performed on realistic $m_\pi=161$ MeV ensembles by the HotQCD and TUMQCD collaboration. From the continuum corrected potential, one may compute in-medium quarkonium spectral functions, which have been used to estimate the $\psi^\prime$ to $J/\psi$ ratio, showing good agreement with the statistical model of hadronization and the most recent measurements by the ALICE collaboration.    To implement the microscopic dynamics of heavy quarkonium based on the complex in-medium potential, the open-quantum-systems approach is promising. Using a clear set of assumptions, one may derive a Lindblad master equation for the reduced density matrix, which to first order leads to unitary time evolution with a stochastic potential. The medium induced noise leads to decoherence of the in-medium quarkonium, which provides an additional mechanism to the dissolution of the in-medium state besides Debye screening. The~implementation of the full Lindblad equation for quarkonium remains work in progress.

\vspace {6pt}
\funding{This work is part of and partially supported by the
  DFG Collaborative Research Centre ``SFB 1225~(ISOQUANT)''.}

\conflictsofinterest{The author declares no conflict of interest.} 



\reftitle{References}


\begin{thebibliography}{999}

\bibitem{Adamczyk:2016dzv} 
Adamczyk, L.. et al. [STAR Collaboration],
$\Upsilon$ production in U + U collisions at $\sqrt{{s}_{NN}}=$ 193 GeV measured with the STAR experiment, \emph{Phys.\ Rev.\ C} {\bf 2016}, \emph{94}, 064904.



\bibitem{Sirunyan:2017lzi} 
Sirunyan, A.M. et al. [CMS Collaboration], Suppression of Excited $\Upsilon$ States Relative to the Ground State in Pb-Pb Collisions at $\sqrt{s_\mathrm{NN}}$=5.02 TeV, \emph{Phys.\ Rev.\ Lett.} {\bf 2018}, \emph{120}, 142301.


\bibitem{Krouppa:2017jlg} 
Krouppa, B.; Rothkopf, A.; Strickland, M., Bottomonium suppression using a lattice QCD vetted potential, \emph{Phys.\ Rev.\ D} {\bf 2018}, \emph{97}, 016017.


\bibitem{Acharya:2018pjd} 
Acharya, S. et al. [ALICE Collaboration], Study of J/$\psi$ azimuthal anisotropy at forward rapidity in Pb-Pb collisions at $ \sqrt{s_{\mathrm{NN}}}=5.02 $ TeV, \emph{ J. High Energy Phys.} {\bf 2019}, \emph{2019}, 012.



\bibitem{Aarts:2014cda} 
Aarts, G.; Allton, C.; Harris, T.; Kim, S.; Lombardo, M.P.; Ryan, S.M.; Skullerud, J.I., The bottomonium spectrum at finite temperature from N$_{f}$ = 2 + 1 lattice QCD, \emph{ J. High Energy Phys.} {\bf 2014}, \emph{2014},~097.


\bibitem{Kim:2018yhk} 
Kim, S.; Petreczky, P.; Rothkopf, A., Quarkonium in-medium properties from realistic lattice NRQCD, \emph{ J. High Energy Phys.} {\bf 2018}, \emph{2018}, 088.


\bibitem{Ding:2012sp} 
Ding, H.T.; Francis, A.; Kaczmarek, O.; Karsch, F.; Satz, H.; Soeldner, W., Charmonium properties in hot quenched lattice QCD, \emph{Phys.\ Rev.\ D} {\bf 2012}, \emph{86}, 014509.


\bibitem{Kelly:2018hsi} 
Kelly, A.; Rothkopf, A.; Skullerud, J.I., Bayesian study of relativistic open and hidden charm in anisotropic lattice QCD, \emph{Phys.\ Rev.\ D} {\bf 2018}, \emph{97}, 114509.


\bibitem{Koma:2007jq} 
Koma, Y.; Koma, M.; Wittig, H., Relativistic corrections to the static potential at O(1/m) and O(1/m**2), \emph{PoS LATTICE} {\bf 2007}, \emph{2007}, 111.


\bibitem{Brambilla:2004jw} 
Brambilla, N.; Pineda, A.; Soto, J.; Vairo, A., Effective field theories for heavy quarkonium, \emph{Rev.\ Mod.\ Phys.} {\bf 2005}, \emph{77}, 1423.

\bibitem{Burnier:2012az} 
Burnier, Y.; Rothkopf, A. , Disentangling the timescales behind the non-perturbative heavy quark potential, \emph{Phys.\ Rev.\ D} {\bf 2012}, \emph{86}, 051503.

\bibitem{Laine:2006ns} 
Laine, M.; Philipsen, O.; Romatschke, P.; Tassler, M., Real-time static potential in hot QCD, \emph{ J. High Energy Phys. } {\bf 2007}, \emph{2007}, 054.

\bibitem{Beraudo:2007ky}  
Beraudo, A.; Blaizot, J.-P.; Ratti, C. , Real and imaginary-time Q anti-Q correlators in a thermal medium, \emph{Nucl.\ Phys.\ A} {\bf 2008}, \emph{806}, 312.

\bibitem{Brambilla:2008cx} 
Brambilla, N.; Ghiglieri, J.; Vairo, A.; Petreczky, P., Static quark-antiquark pairs at finite temperature, \emph{Phys.\ Rev.\ D} {\bf 2008}, \emph{78}, 014017.

\bibitem{Akamatsu:2011se} 
Akamatsu, Y.; Rothkopf, A., Stochastic potential and quantum decoherence of heavy quarkonium in the quark-gluon plasma, \emph{Phys.\ Rev.\ D} {\bf 2012}, \emph{85}, 105011.

\bibitem{Rothkopf:2009pk} 
Rothkopf, A.; Hatsuda, T.; Sasaki, S., Proper heavy-quark potential from a spectral decomposition of the thermal Wilson loop, \emph{PoS LAT} {\bf 2009}, \emph{2009}, 162.

\bibitem{Rothkopf:2011db} 
Rothkopf, A.; Hatsuda, T.; Sasaki, S. , Complex Heavy-Quark Potential at Finite Temperature from Lattice QCD, \emph{Phys.\ Rev.\ Lett.} {\bf 2012}, \emph{108}, 162001.

\bibitem{Rothkopf:2019dzu} 
Rothkopf, A., Bayesian techniques and applications to QCD, \emph{arXiv} \textbf{2019},
arXiv:1903.02293.

\bibitem{Tripolt:2018xeo} 
Tripolt, R.A.; Gubler, P.; Ulybyshev, M.; von Smekal, L., Numerical analytic continuation of Euclidean data, \emph{Comput.\ Phys.\ Commun.} {\bf 2019}, \emph{237}, 129.

\bibitem{Burnier:2013nla}  
Burnier, Y.; Rothkopf, A. , Bayesian Approach to Spectral Function Reconstruction for Euclidean Quantum Field Theories, \emph{Phys.\ Rev.\ Lett.} {\bf 2013}, \emph{111}, 182003.


\bibitem{Petreczky:2017aiz} 
TUMQCD Collaboration, Lattice Calculations of Heavy Quark Potential at Finite Temperature, \emph{Nucl.\ Phys.\ A} {\bf 2017}, 967, 592.


\bibitem{Bazavov:2017dsy} 
Bazavov, A.; Petreczky, P.; Weber, J.H., Equation of State in 2+1 Flavor QCD at High Temperatures, \emph{Phys.\ Rev.\ D} {\bf 2018}, \emph{97},  014510.


\bibitem{Bazavov:2014pvz} 
HotQCD Collaboration, Equation of state in ( 2+1 )-flavor QCD, \emph{Phys.\ Rev.\ D} {\bf 2014}, \emph{90}, 094503.


\bibitem{Petreczky:2018xuh} 
Petreczky, P.; Rothkopf, A.; Weber, J., Realistic in-medium heavy-quark potential from high statistics lattice QCD simulations, \emph{Nucl.\ Phys.\ A} {\bf 2019}, \emph{982}, 735.


\bibitem{Burnier:2015tda} 
Burnier, Y.; Kaczmarek, O.; Rothkopf, A., Quarkonium at finite temperature: Towards realistic phenomenology from first principles, \emph{ J. High Energy Phys.} {\bf 2015}, \emph{2015}, 101.

\bibitem{Lafferty:2019jpr} 
 Lafferty, D. and Rothkopf, A., Improved Gauss-law model and in-medium heavy quarkonium at finite density and velocity, arXiv:1906.00035 [hep-ph].

\bibitem{LaffertyZimanyi}
Lafferty, D. and Rothkopf, A. , Quarkonium Phenomenology from a Generalised Gauss Law, 
Universe {\bf 5}, no. 5, 119 (2019)


\bibitem{Kajimoto:2017rel} 
Kajimoto, S.; Akamatsu, Y.; Asakawa, M.; Rothkopf, A., Dynamical dissociation of quarkonia by wave function decoherence, \emph{Phys.\ Rev.\ D} {\bf 2018}, \emph{97}, 014003.



\bibitem{Andronic:2009sv} 
Andronic, A.; Beutler, F.; Braun-Munzinger, P.; Redlich, K.; Stachel, J., Statistical hadronization of heavy flavor quarks in elementary collisions: Successes and failures, \emph{Phys.\ Lett.\ B} {\bf 2009}, \emph{678}, 350.


\bibitem{Andronic:2018vqh} 
Andronic, A.; Braun-Munzinger, P.; Köhler, M.K.; Stachel, J., Testing charm quark thermalisation within the Statistical Hadronisation Model, \emph{Nucl.\ Phys.\ A} {\bf 2019}, \emph{982}, 759.


\bibitem{Akamatsu:2012vt} 
Akamatsu, Y., Real-time quantum dynamics of heavy quark systems at high temperature, \emph{Phys.\ Rev.\ D} {\bf 2013}, \emph{87}, 045016.


\bibitem{Akamatsu:2014qsa} 
Akamatsu, Y., Heavy quark master equations in the Lindblad form at high temperatures, \emph{Phys.\ Rev.\ D} {\bf 2015}, \emph{91}, 056002.


\bibitem{DeBoni:2017ocl} 
De Boni, D., Fate of in-medium heavy quarks via a Lindblad equation, \emph{ J. High Energy Phys.} {\bf 2017}, \emph{2017}, 064.


\bibitem{Blaizot:2017ypk} 
Blaizot, J.P.; Escobedo, M.A., Quantum and classical dynamics of heavy quarks in a quark-gluon plasma, \emph{ J. High Energy Phys.} {\bf 2018}, \emph{2018}, 034.


\bibitem{Brambilla:2017zei} 
Brambilla, N.; Escobedo, M.A.; Soto, J.; Vairo, A., Heavy quarkonium suppression in a fireball, \emph{Phys.\ Rev.\ D} {\bf 2018}, \emph{97}, 074009.

\bibitem{Yao:2018nmy} 
  X.~Yao and T.~Mehen, Quarkonium in-Medium Transport Equation Derived from First Principles, \emph{arXiv} \textbf{2019}, arXiv:1811.07027.

\bibitem{Brambilla:2019tpt} 
Brambilla, N.; Escobedo, M.A.; Vairo, A.; Griend, P.V, Transport coefficients from in medium quarkonium dynamics, \emph{arXiv} \textbf{2019}, 
arXiv:1903.08063.



\bibitem{Akamatsu:2018xim} 
Akamatsu, Y.; Asakawa, M.; Kajimoto, S.; Rothkopf, A., Quantum dissipation of a heavy quark from a nonlinear stochastic Schr\"odinger equation, \emph{ J. High Energy Phys.} {\bf 2018}, \emph{2018}, 029.
\end{thebibliography}



\end{document}